# Why Task-Based Training is Superior to Traditional Training Methods

Kath McGuire
Small Spark
St John's Innovation Centre, Cowley Road, Cambridge, CB4 0WS
kath@smallspark.co.uk

**ABSTRACT**

*The risks of spreadsheet use do not just come from the misuse of formulae. As such, training needs to go beyond this technical aspect of spreadsheet use and look at the spreadsheet in its full business context. While standard training is by and large unable to do this, task-based training is perfectly suited to a contextual approach to training.*

## 1      THE SITUATION

### 1.1    What Are The Risks Of Spreadsheet Use?

The main risks associated with spreadsheet use are the prevalence of errors (which can cause inadvertent error and make fraud harder to detect) and the gross misuse of time by spreadsheet operators (designers, users, etc.).

The effects of spreadsheet errors are well documented: bad decision making, non-compliance, ease of fraud etc. "The biggest fear for organisations is losing data or suffering losses due to unchecked inaccurate data." [Baxter, 2007] It has been reported that as many as 90% of spreadsheets contain errors and that many of those errors can be critical to business [Bewig, 2005]. The EuSpRIG website [EuSpRIG] contains many reports of specific instances of business loss due to spreadsheet error.

The effects of time-wasting are not as well documented, as they hit businesses in a more indirect manner. However, staff productivity is definitely affected by fighting an unfriendly system and by wasting hours doing things a long way. This waste can cost money.

### 1.2    What Are The Causes Of These Risks?

There are two levels at which spreadsheet errors occur: the user level and the business level.

**User Error**

We all make mistakes regardless of whether we are designing, auditing or using a spreadsheet system. While natural human error cannot be eradicated by training, an awareness of possible errors can assist with detection and remedy [Purser, Chadwick, 2006].

Lack of technical skills (e.g.: choosing the wrong 4th argument in a vlookup) and lack of awareness of software features (e.g.: cutting and pasting swathes of data rather than using lookup functions) can be addressed by training.

**Business Error**

For most business people – wherever placed in the hierarchy – there is a lack of awareness of what their colleagues actually do. Most users do not know where the information in their spreadsheet comes from or goes to. This can lead to errors in the boundaries between work spaces, roles or tasks. Examples include using misclassifications of products, users repeating calculations because they are unaware that another has the summary they actually need, etc.

Many of these business errors are logical errors or errors of omission, which Panko has stated are more difficult to detect and potentially more dangerous than the mechanical errors characterised by user error. However, all three are damaging and prevalent [Panko, 2005].

## 2 THE SOLUTION

We believe that task-based training is required for most effective management of spreadsheet systems in any organisation.

### 1.3 What Is Task-Based Training?

Task-based training looks at the task that needs to be done, rather than looking in isolation at the way the software works.

Task-based trainers have experience in the way spreadsheets are actually used in the real world rather than a theoretical idea of how they should be used. They combine this experience with knowledge of the client's circumstances.

The task is put in the context of the business processes that surround it. For example, people are not taught a conditional sum as a spreadsheet feature, they are taught how to report sales activity by client for the past month – it may be that the best way to do that is a conditional sum, a pivot table, subtotals, a macro or perhaps even using a database. That decision is based on factors such as the nature of the organisation, the source-data, the output requirements and the user.

Much of the training that Small Spark conducts is task-based, though we do also engage in standard training. We have talked to our clients and some of their feedback has been included in this paper.

### 1.4 What Are The Alternatives To Task-Based Training?

Task-based training is not the only type of training and is far from the most popular.

Off-the-shelf training packages are the most common form of training. They usually divide training provision broadly by perceived skill level. They are often package and version specific. As they frequently include people with different uses of spreadsheets, the training must be general and is therefore unable to adapt to the particular needs of the trainees.

While this is very common, most spreadsheet users have undergone no formal training whatsoever. Many learn on the job from their predecessor or colleagues and some are self-taught.

**1.5  Why Is Task-Based Training Superior?**

By the end of a task-based training session, trainees will have completed tasks, have set up systems for future tasks, have seen some of their problems solved and have worked on actual spreadsheets. This makes task-based training more effective. The trainee is learning both the mechanical skills associated with the task (the conditional sum, the data pilot, etc.) as well as the reason for use of this particular skill in a given instance. As such the skills being taught can be applied not only in the current circumstance but in further spreadsheet work involving different tasks.

Since task-based training by its very definition takes account of the business processes in which the users finds themselves, a task-based trainer is able to be more effective than a standard trainer of a comparable skill level. When we have engaged in standard training we have received feedback which highlights the drawbacks of standard training in comparison to task-based training: *"some a-ha moments that I will find very helpful and then other parts which just seemed a long way of doing things I already do"*. In a task-based training session it is possible (and in fact essential) to find out if a solution is appropriate for the trainee and then train them in what is appropriate for them.

When one of our clients was asked about her attitude to off-the-shelf training as opposed to task-based training she replied: *"if I went to one of their half-day courses or evening classes, they may not cover exactly what I'm looking for"*.

Task-based training can help to ensure that the right spreadsheet solution is implemented appropriately by the right people. Knowing about the function is not enough – it is important to know when to use it: *"We covered things that I always knew were there but never knew how to access"*.

Also, like other training provision, task-based training can help with user errors such as incorrect use of functions. It is possible to build the user's confidence by fully explaining the best solution and how to implement it appropriately for their task. One of our clients has said about her staff *"They know they're doing the right thing and I think it makes them more productive"*.

**Spreadsheets In Context**

In order to design and implement an effective spreadsheet solution, it is important to have business knowledge to ensure that data is turned into useful information, and technical knowledge to turn the data into information in a robust and efficient manner.

The business knowledge exists in-house. Technical knowledge is sought through training. If this technical knowledge does not take account of the business knowledge, the solution will not be optimal. Standard training seeks to provide technical skills within a vacuum. Task-based training takes account of the business knowledge to ensure that the appropriate technical skills are imparted.

A client had an Excel-based data manipulation task. She was new to the organisation and had been given very little induction. Her boss knew what needed to be done (business knowledge) but did not know how to do it. The technical staff could have created an algorithm for the data processing but could not have implemented it in Excel. Standard training is at a loss in such situations as the trainee does not know which features would be useful. Task-based training is vital because the business knowledge can be incorporated into the technical spreadsheet training. *"You took it apart and removed the mystique from it ... It's a nice routine. It takes seconds now."*

Business knowledge provides information about the business context of the spreadsheet, the skills of the users, as well as hardware and software considerations of the business. All of these affect the design, use and improvement of spreadsheets. Business knowledge is important to providing relevant training: *"you understand my business and that makes a real difference to the training"*.

*Business context*

The business context of the spreadsheet includes how information flows through the organisation and who needs to access the information. This can have an impact on whether existing spreadsheets are rewritten or streamlined.

The correct tools will vary depending on the size and scope of the solution. While pasting data is usually a bad idea – and standard training should warn against it – it may be that it is the best solution for a particular system. Task-based training allows for this and the trainer can explain the relevant dangers and suggest appropriate techniques to mitigate the risk (such as check sums).

*Skill awareness*

Training a few key employees in VBA or other 'advanced' techniques – pivot tables in Excel, the data pilot in OpenOffice, etc. – is often popular. But if their colleagues cannot use these features, this knowledge is not adding value to the organisation. Using lookups to generate many static tables might be more suitable if this reflects the skill level of all users. This is a decision that standard training cannot make, and that task-based training must make.

It is important that the spreadsheet solution being implemented is sufficiently user-friendly and allows for all the actions a user might want to take, so as to remove the need for hacking around it. This sort of design consideration is much easier to teach within a task-based training framework that takes into account the requirements and personalities of all users.

*Software and Hardware*

Not all spreadsheet users use the same packages or versions. This variation of software is found within organisations, not just between them. Task-based training is better able to appreciate differences in software between users, or between machines used by the same user.

Hardware and version control systems (if any exist) can also affect the size, speed and scope of the right solution. A standard training provider is unable to train with this in mind whereas it is information that a task-based trainer does have access to.

**Relevant Error Management**

Teaching people what spreadsheets can do is the most important part of training, but it is also vital to teach people what they cannot do. The limitations and quirks of solutions are often not explained in training because the nature of these limitations may only become apparent when the solution is used in a particular context.

Task-based training can pre-empt many of these limitations and can make allowances for them. Error checks can be built in, spreadsheets can be commented and staff can be taught what sort of errors to look for and how to spot them.

The nature of task-based training means that external eyes are looking at the solution and can spot actual or potential errors that may elude the existing users.

**Documentation**

Documentation of spreadsheets is important and unusual. When documentation is created, it is frequently consulted initially but then filed away and forgotten.

Task-based training discusses processes. It questions why things are done the way they are done and brings to the forefront the reasons behind the processes. When notes are provided as part of a task-based training session, they can become a part of the documentation of the solution.

## 3 THE CONCLUSION

The problems with spreadsheet use go beyond teaching someone how to use a particular spreadsheet function. The right skills to teach and the right solution to design can be appreciated most effectively through task-based training provision.

Task-based training is better able than standard training provision to effectively mitigate the drawbacks of spreadsheet use for ALL spreadsheet users (regardless of skill level, seniority, qualifications, job description, etc.) in ALL organisations (regardless of size, complexity, training budget, etc.).

Blank Page